\documentclass[10pt,journal,twoside,print]{ieeecolor}
\usepackage{generic}

\usepackage{etoolbox}
\makeatletter
\@ifundefined{color@begingroup}%
  {\let\color@begingroup\relax
   \let\color@endgroup\relax}{}%
\def\fix@ieeecolor@hbox#1{%
  \hbox{\color@begingroup#1\color@endgroup}}
\patchcmd\@makecaption{\hbox}{\fix@ieeecolor@hbox}{}{\FAILED}
\patchcmd\@makecaption{\hbox}{\fix@ieeecolor@hbox}{}{\FAILED}

\def\BibTeX{{\rm B\kern-.05em{\sc i\kern-.025em b}\kern-.08em
    T\kern-.1667em\lower.7ex\hbox{E}\kern-.125emX}}
\usepackage{graphicx,epstopdf,multirow,booktabs}
\usepackage{amsmath,amssymb,amsthm, bm}
\usepackage{cases}
\usepackage[T1]{fontenc}
\usepackage{hyperref}
\allowdisplaybreaks
\hypersetup{
     colorlinks =true,
     linkcolor = blue,
     filecolor = blue,
     citecolor = magenta,      
     urlcolor = magenta,
     }
\usepackage{setspace}
\usepackage{siunitx}
\usepackage{csvsimple}
\usepackage{url}
\usepackage[T1]{fontenc}

\let\labelindent\relax
\usepackage{enumitem, xspace}
\makeatletter
\def\ps@headings{%
  \def\@oddhead{%
    \vbox{%
      \hbox to \textwidth{%
        \scriptsize\sffamily
        \rightmark\hfill\thepage
      }%
      \vskip 2pt
      {\color{subsectioncolor}\hrule height 1pt}%
    }%
  }%
  \def\@evenhead{%
    \vbox{%
      \hbox to \textwidth{%
        \scriptsize\sffamily
        \thepage\hfill\leftmark
      }%
      \vskip 2pt
      {\color{subsectioncolor}\hrule height 1pt}%
    }%
  }%
  \def\@oddfoot{}%
  \def\@evenfoot{}%
}
\makeatother
\newtheorem{proposition}{Proposition}

\newtheorem{remark}{Remark}
\newtheorem{corollary}{Corollary}

\begin{document}
\title{Controller- and Communication-Aware Flexibility Margins for Electric Distribution Networks}
\author{Saif R. Kazi and Kaarthik Sundar
\thanks{The authors are affiliated with Los Alamos National Laboratory, Los Alamos, NM. Corresponding author: \texttt{kaarthik@lanl.gov}}
}

\maketitle
\begin{abstract}
This paper develops a framework for quantifying operational flexibility in radial electric distribution systems under coupled active- and reactive-load uncertainty. Flexibility is defined as the largest scaling of a prescribed uncertainty set that can be accommodated while satisfying network and actuator limits under a given control architecture. We consider a one-sided Euclidean set for robust demand-growth capability, with a direct extension to bidirectional variations. Under LinDistFlow, controllable distributed resources and feeder-head exchange with the upstream grid jointly balance active-power disturbances, while local Volt/VAR feedback regulates reactive power. The resulting affine closed-loop disturbance response admits exact second-order-cone robust counterparts. For fixed Volt/VAR gains, a convex reformulation jointly optimizes the flexibility margin and active-power participation factors, including feeder-head participation. A sparse load-to-generator feedforward map further captures communication constraints. Numerical studies on the Iowa-240 feeder demonstrate how amount of control authority and communication structure shape the achievable flexibility margin.
\end{abstract}

\begin{IEEEkeywords}
Distribution systems, operational flexibility, networked control systems, actuator limits, Volt/VAR control, communication constraints
\end{IEEEkeywords}

\section{Introduction} \label{sec:intro}
\IEEEPARstart{E}{lectric} distribution systems are increasingly populated by photovoltaic generation, battery storage, electric vehicles, flexible loads, and other distributed energy resources (DERs). At the same time,
electrification and variability in net demand are creating a wider range of operating conditions and power-flow patterns. These changes introduce
uncertainty, but also new control authority: appropriately coordinated DERs can provide active- and reactive-power flexibility \cite{riaz2021modelling}.

For planners and operators, the relevant question is therefore not only whether a feeder can accommodate a forecast operating point or new interconnection, but also how much additional disturbance it can tolerate and how effectively it can use available control resources to accommodate it. A quantitative measure of this operational headroom enables comparison of DER deployments, actuator capabilities, control architectures, and communication structures, and can inform when operational flexibility may defer network upgrades.

Related work characterizes DER flexibility through feasible active- and reactive-power regions \cite{riaz2021modelling,kundu2018approximating}, and studies hosting capacity and approximations of power-flow feasibility regions \cite{nazir2019convex,lee2019convex,christianen2025polyhedral}. Stochastic and robust optimal power flow address operation under prescribed uncertainty models \cite{bienstock2014chance,lee2021robust,roald2023power,agarwal2022employing}, while affine decision rules provide tractable robust-control policies
\cite{ben2004adjustable,goulart2006optimization}. Distance-to-infeasibility and adversarial-perturbation methods instead quantify proximity to feasibility boundaries \cite{jeeninga2024distance,chevalier2026identifying,shan2026voltage}.
\textit{Here, we ask a complementary closed-loop question: for a given feeder, nominal operating condition, set of actuators, and control and information architecture, how large can coupled active- and reactive-load disturbances become while all network and actuator constraints remain satisfied?}

A key aspect of this question is how active-power disturbances are balanced. Since the feeder is connected to an upstream network, incremental active-power demand need not be supplied entirely by controllable DERs within the feeder. Instead, downstream DERs and the upstream grid may supply it jointly through feeder-head exchange. We model this sharing explicitly through participation factors and impose limits on feeder-head interchange. We then define operational flexibility as the largest scaling of a prescribed uncertainty set contained in the closed-loop feasible disturbance region. A one-sided Euclidean set quantifies robust demand-growth capability, while a symmetric set gives the corresponding bidirectional margin.

Under LinDistFlow \cite{baran1989optimal} approximation of the physics of power flow in distribution networks, the active-power participation policy and local Volt/VAR feedback make the closed-loop feeder response affine in the load disturbance. Consequently, each network and actuator limit can be expressed as an affine inequality in the disturbance. Requiring such a limit to hold for every disturbance in the prescribed uncertainty set defines a robust constraint; for the Euclidean uncertainty sets considered here, this semi-infinite requirement can be replaced exactly by an equivalent deterministic norm inequality. These deterministic equivalents, referred to as robust counterparts, yield tractable formulations for certifying the largest admissible uncertainty-set radius and, for fixed Volt/VAR gains, for optimizing the active-power participation policy. The resulting flexibility margin therefore reflects not only physical network headroom, but also the control authority and information structure available to access it. Here, exactness is with respect to the adopted LinDistFlow model and engineering constraints, rather than the non-linear DistFlow equations \cite{baran1989optimal}.

The contributions are threefold. First, we formulate operational flexibility as a closed-loop uncertainty-accommodation margin that depends explicitly on the feeder, downstream actuators, feeder-head balancing authority, controller, and information architecture. Second, for fixed Volt/VAR gains, we derive exact robust formulations; for fixed active-power participation factors, the margin comes from a linear program in a single scalar variable, while jointly optimizing the margin and active-power participation factors admits an exact second-order cone programming (SOCP) reformulation. Third, we generalize global participation factors to a sparse load-to-generator feedforward map, enabling communication constraints while preserving convexity.

The remainder of the paper is organized as follows. Section \ref{sec:modeling} develops the feeder and controller model, Section \ref{sec:robust} derives the flexibility-certification formulations, Section \ref{sec:comm} introduces communication-constrained feedforward control, and Section \ref{sec:results} presents computational results followed by concluding remarks in Section \ref{sec:conclusion}. The Appendix provides supporting proofs.

\section{Modeling} \label{sec:modeling}
We start this section by presenting the model of a radial distribution network as an algebraic plant. 
\subsection{Radial distribution feeder as an algebraic plant} \label{subsec:open-loop}
Consider a radial distribution network with root-bus $0$, the set of non-root-buses $\mathcal N$, and the set of distribution lines $\mathcal E$. Let $n \triangleq |\mathcal N|$. For each non-root-bus $j \in \mathcal N$, let $\pi(j)$ denote its parent bus and $\mathcal C(j)$ denote its set of children. Associate with each line $(\pi(j),j) \in \mathcal E$ the active- and reactive-power flows $p_j$ and $q_j$, resistance $r_j$, and reactance $x_j$. Let $v_j$ denote the squared voltage magnitude at bus $j \in \mathcal N \cup \{0\}$, with the root-bus voltage fixed at $v_0=\bar v_0$.
Under the LinDistFlow approximation, the nodal real- and reactive-power-balance and branch voltage-drop equations are
\begin{subequations}
\begin{gather}
    p_j - \sum_{k \in \mathcal C(j)} p_k = p_j^d - p_j^g \label{eq:real-balance} \\ 
    q_j - \sum_{k \in \mathcal C(j)} q_k = q_j^d - q_j^g \label{eq:reactive-balance} \\ 
    v_{j} - v_{\pi(j)} + 2 \cdot ( r_j  p_j + x_j q_j) = 0. \label{eq:voltage-drop}
\end{gather}
\label{eq:lindistflow}
\end{subequations}
where $p_j^{d}$ and $q_j^{d}$ denote active and reactive demand, while $p_j^{g}$ and $q_j^{g}$ denote controllable active and reactive generation. 

From a control-theoretic perspective, we interpret the power-flow variables as the \textit{algebraic state} of the feeder. Define the stacked branch-flow and voltage vectors 
\begin{gather*}
    \bm p \triangleq \operatorname{col}(p_j)_{j \in \mathcal N}, ~\bm q \triangleq \operatorname{col}(q_j)_{j \in \mathcal N}, \text{ and } \bm v \triangleq \operatorname{col}(v_j)_{j \in \mathcal N}
\end{gather*}
We additionally retain the active- and reactive-power exchanges with the upstream network at the feeder head, denoted by $p_0^{\mathrm{sl}}$ and $q_0^{\mathrm{sl}}$, respectively. The algebraic state is therefore
\begin{gather*}
    \bm z \triangleq \operatorname{col}(\bm p,\bm q, \bm v, p_0^{\mathrm{sl}}, q_0^{\mathrm{sl}}) \in \mathbb R^{3n + 2}.
\end{gather*}
The term ``state'' is used here in the algebraic sense. The LinDistFlow model is quasi-steady-state and contains no temporal state evolution. Although $p_0^{\mathrm{sl}}$ and $q_0^{\mathrm{sl}}$ are algebraically determined by the branch flows, we retain them explicitly because feeder-head interchange limits will later be imposed as engineering constraints. We also define the controllable injection vector $\bm u$ and the exogenous input vector $\bm d$ as
\begin{gather*}
    \bm u \triangleq \operatorname{col}(\bm p^g, \bm q^g), \qquad  \bm d \triangleq \operatorname{col}(\bm p^d, \bm q^d).
\end{gather*}
Here, $\bm p^g$ and $\bm q^g$ are the vectors of nodal active and reactive injections, and $\bm p^d $ and $\bm q^d$ are the vectors of active and reactive demands. If controllable injections are present only at a subset of buses, the corresponding entries may either be fixed to zero or represented through suitable selection matrices. In the subsequent control-theoretic formulation, we parameterize variations in $\bm d$ about a nominal operating condition with a disturbance variable. 

To obtain a compact matrix representation, define the branch-balance matrix $F\in\mathbb R^{n\times n}$ such that
\begin{gather*}
    (F \bm p)_j = p_j - \sum_{k\in \mathcal C(j)} p_k.
\end{gather*}
Let 
    $R\triangleq \operatorname{diag}(r_j),~ X\triangleq \operatorname{diag}(x_j)$.
We also observe that 
\begin{gather*}
    (F^\intercal \bm v)_j \triangleq 
    \begin{cases} 
        v_j-v_{\pi(j)} & \text{ when } \pi(j)\neq 0 \\ 
        v_j            & \text{ when } \pi(j) = 0.
    \end{cases}
\end{gather*}
Finally, define $\bm s\in\{0, 1\}^n$ such that $s_j=1$ only when $\pi(j)=0$. The voltage equations can then be written compactly as
\begin{gather*}
    2 R \bm p + 2 X \bm q + F^\intercal  \bm v = \bm s \bar{v}_0.
\end{gather*}
With the convention that positive $p_0^{\mathrm{sl}}$ and  $q_0^{\mathrm{sl}}$ denote real- and reactive-power supplied from the root into the feeder, $p_0^{\mathrm{sl}} = \bm s^\intercal \bm p$ and $q_0^{\mathrm{sl}} = \bm s^\intercal \bm q$. The complete open-loop feeder model can therefore be written as the static algebraic plant
\begin{gather}
    A \bm z = B_d \bm d + B_u \bm u + \bm b_0 \label{eq:open-loop}
\end{gather}
where
\begin{gather*}
    A \triangleq \begin{bmatrix} 
    F & 0 & 0 & 0 & 0\\ 
    0 & F & 0 & 0 & 0\\ 
    2R & 2X & F^\intercal & 0 & 0\\
    -\bm s^\intercal & 0 & 0 & 1 & 0\\
    0 &-\bm s^\intercal & 0 & 0 & 1
    \end{bmatrix}, ~
    B_d = -B_u \triangleq \begin{bmatrix}
        I & 0 \\ 
        0 & I \\ 
        0 & 0 \\ 
        0 & 0 \\ 
        0 & 0
    \end{bmatrix}
\end{gather*}
and $\bm b_0 \triangleq \operatorname{col}(0, 0, \bm s\bar{v}_0, 0, 0)$. As shown in Corollary \ref{cor:A-openloop}, $A$ is non-singular for a radial feeder, and hence the feeder may equivalently be expressed as the affine input-state map
\begin{equation*}
\begin{gathered}
    \bm z = G_d \bm d + G_u \bm u + \bm g_0 \\ 
    \text{ where }G_d \triangleq A^{-1} B_d,~ G_u \triangleq A^{-1} B_u,~ \bm g_0 \triangleq  A^{-1}\bm b_0
\end{gathered}
\end{equation*}
This affine representation separates the effects of exogenous demand and controllable generation and provides the basis for the closed-loop model developed in the subsequent sections.

\subsection{Local static Volt/VAR output feedback} \label{subsec:outputfeedback}
We next introduce the reactive-power control channel through a local static Volt/VAR law expressed in squared-voltage coordinates. Let $\mathcal N_I \subseteq \mathcal N$ denote the set of buses equipped with controllable inverter-based reactive-power resources. At each inverter bus $j\in\mathcal N_I$, the reactive-power injection is adjusted in response to the locally measured squared voltage magnitude according to the Volt/VAR law
\begin{gather}
    q_j^g = \bar q_j^g - k_v (v_j - v_j^{\mathrm{ref}}) \label{eq:local-vv}
\end{gather}
where $\bar q_j^g$ is the reactive-power set-point at zero voltage error, $v_j^{\mathrm{ref}}$ is the reference squared voltage magnitude, and $k_j\geqslant0$ is the corresponding Volt/VAR gain. Thus, $\bar q_j^g$ need not equal the realized reactive-power injection at the nominal closed-loop operating point unless the nominal voltage equals $v_j^{\mathrm{ref}}$. At buses without voltage-responsive reactive-power control, we set $k_j=0$; if no
reactive-power resource is present, we additionally set
$\bar q_j^g=0$. Reactive-power capability limits are imposed
separately as engineering constraints. Now define
\begin{gather*}
    K_v \triangleq \operatorname{diag}(k_1, k_2, \dots, k_n) \succeq 0
\end{gather*}
with non-zero entries only at buses in $\mathcal N_I$. Stacking the bus-level control laws results in
\begin{gather}
    \bm q^g = \bar {\bm q}^g - K_v \bm v + K_v \bm v^{\mathrm{ref}}. \label{eq:local-vv-vector}
\end{gather}
Substituting the local Volt/VAR control law \eqref{eq:local-vv-vector} into the reactive flow balance equation and collecting the state-dependent terms to the left-hand side, we obtain 
\begin{gather*}
    F\bm q - K_v \bm v = \bm q^d - \bar{\bm q}^g - K_v \bm v^{\mathrm{ref}}.
\end{gather*}
Hence, $\bm q^g$ is eliminated as an independent input, while the term $-K_v\bm v$ enters the algebraic operator. Local Volt/VAR control therefore acts as static output feedback on the feeder. Accordingly, define the closed-loop algebraic operator and affine term as
\begin{equation}
\begin{gathered}
    A_{\mathrm{cl}}(K_v) \triangleq \begin{bmatrix} 
    F & 0 & 0 & 0 & 0\\ 
    0 & F & -K_v & 0 & 0\\ 
    2R & 2X & F^\intercal & 0 & 0\\
    -\bm s^\intercal & 0 & 0 & 1 & 0 \\
    0 &-\bm s^\intercal & 0 & 0 & 1
    \end{bmatrix}, \\ 
    \bm b_{\mathrm{cl}}(K_v) \triangleq \operatorname{col}(0,  -\bar{\bm q}^g - K_v \bm v^{\mathrm{ref}}, \bm s \bar v_0, 0, 0).
\end{gathered}\label{eq:closed-loop-matrices}
\end{equation}
Since the controller closes a static algebraic feedback loop, a unique closed-loop solution requires $A_{\mathrm{cl}}(K_v)$ to be nonsingular. 
\begin{proposition}[Closed-loop well-posedness] \label{prop:Acl} For $X\succeq 0$ and $K_v\succeq 0$, $A_{\mathrm{cl}}(K_v)$ is nonsingular. 
\end{proposition}
\begin{proof}
    See Appendix \ref{subsec:app1}. 
\end{proof}
Proposition~\ref{prop:Acl} establishes this property for $K_v, X\succeq0$ under the adopted feeder model. Using \eqref{eq:closed-loop-matrices}, the feeder model can be expressed as 
\begin{gather}
    A_{\mathrm{cl}}(K_v) \bm z = B_d \bm d - B_p \bm p^g  + \bm b_{\mathrm{cl}}(K_v) \label{eq:after-fb}
\end{gather}
where $B_p \triangleq [I, 0, 0, 0, 0]^\intercal$.Thus, local Volt/VAR control is embedded in the feeder model as
static output feedback, while $\bm p^g$ remains an explicit
active-power input whose disturbance-dependent balancing policy is introduced next.

\subsection{Load disturbances and active-power balancing} \label{subsec:dist-ff}
We model coupled active- and reactive-load variations by $\bm \eta \in \mathbb R^n$, where $\eta_j$ denotes the fractional variation of the load at bus $j$. Specifically, 
\begin{gather}
    p_j^d(\eta_j) = \bar p_j^d(1+\eta_j) \text{ and } q_j^d(\eta_j) = \bar q_j^d(1 + \eta_j) \label{eq:d-model}
\end{gather}
so that the nominal power factor at each bus is preserved. In vector form, 
\begin{gather*}
    \bm p^d(\bm \eta) = \bm {\bar p}^d + D_p \bm \eta \text{ and } \bm q^d(\bm \eta) = \bar {\bm q}^d + D_q \bm \eta 
\end{gather*}
where $D_p \triangleq \operatorname{diag}(\bm {\bar p}^d)$ and $D_q \triangleq \operatorname{diag}(\bm {\bar q}^d)$. At buses with zero nominal load, the corresponding entries of $\bm \eta$ are immaterial and may be set to zero without loss of generality. 

We focus primarily on the demand-growth disturbances and adopt the one-sided Euclidean set 
\begin{gather}
    \mathcal U_+(\rho) \triangleq \{ \bm \eta \in \mathbb R^n : \|\bm \eta\|_2 \leqslant \rho, \bm \eta \geqslant \bm 0 \}, \label{eq:eta-ball}
\end{gather}
where $\rho\geqslant 0$ scales the admissible disturbance magnitude. 
\begin{remark}[Interpretation of $\mathcal U_+(\rho)$] \label{rem:interpretaion}
The radius $\rho$ measures the Euclidean magnitude of the bus-wise fractional load variations and should not be interpreted directly as a uniform percentage increase in feeder demand. For instance, if $100$ load buses each experience a $5\%$ increase, then $\eta_j=0.05$ at each bus and $\|\eta\|_2=0.05\sqrt{100}=0.5$, so this disturbance lies on the boundary of $\mathcal U_+(0.5)$. The same set also contains non-uniform spatial growth patterns whose Euclidean norm does not exceed $0.5$. 
\end{remark}
For bidirectional load variations, the corresponding symmetric set  is 
\begin{gather}
    \mathcal U_2(\rho) \triangleq \{ \bm \eta \in \mathbb R^n : \|\bm \eta\|_2 \leqslant \rho\}. \label{eq:eta-ball-euc}
\end{gather}
Unless stated otherwise, we use $\mathcal U(\rho) \equiv \mathcal U_+(\rho)$. 

The active-power disturbance is balanced jointly by the controllable resources within the feeder and by the upstream network. Let $\alpha_0$ denote the fraction of the aggregate active-power imbalance supplied by the upstream network, and let $\alpha_j$ denote the fraction assigned to the participating generator at bus $j$. We now collect the downstream participation factors in the $\bm \alpha \in \mathbb R^n$, with $\alpha_j =0$ at buses without participating generation, and impose 
\begin{gather}
   \alpha_0 \geqslant 0,\qquad
    \bm\alpha \geqslant \bm 0,\qquad
    \alpha_0 + \bm 1^\intercal \bm\alpha = 1, \label{eq:alpha-policy} 
\end{gather}
Thus, $\alpha_0 =0$ corresponds to complete balancing by participating resources within the feeder, whereas $\alpha_0 > 0$ allows the upstream network to supply a portion of the imbalance. 

Under \eqref{eq:d-model}, the aggregate active-power disturbance is 
\begin{gather}
    \Delta p^d(\bm \eta) = \bm 1^\intercal D_p \bm \eta = (\bar {\bm p}^d)^\intercal \bm \eta. \label{eq:aggregate-active-disturbance}
\end{gather}
The downstream active-power response is modeled by the global participation policy 
\begin{gather}
    \bm p^g(\bm \eta) = \bar{\bm p}^g + \bm \alpha (\bar {\bm p}^d)^\intercal \bm \eta = \bar{\bm p}^g + L_p(\bm \alpha) \bm \eta.  \label{eq:active-policy}
\end{gather}
where, $L_p(\bm \alpha) \triangleq \bm \alpha (\bar {\bm p}^d)^\intercal$. Hence, every participating resource responds to the same aggregate active-power disturbance, with $\bm\alpha$ determining how that response is distributed among the resources.

The feeder-head active-power exchange is not an independent control input; rather, it is the residual balancing response implied by the network power balance. From
$\bm1^\intercal F=\bm s^\intercal$ and the active-power balance
equations,
\begin{gather*}
    p_0^{\mathrm{sl}}(\bm \eta) = \bm1^\intercal
    \left( \bm p^d(\bm \eta)-\bm p^g(\bm \eta)\right).
\end{gather*}
Defining the nominal feeder-head exchange as
\begin{gather*}
    \bar p_0^{\mathrm{sl}} \triangleq \bm1^\intercal
    \left( \bar{\bm p}^d-\bar{\bm p}^g \right),
\end{gather*}
and using \eqref{eq:d-model}, \eqref{eq:alpha-policy} and \eqref{eq:active-policy} yields
\begin{align}
    p_0^{\mathrm{sl}}(\eta) &= \bar p_0^{\mathrm{sl}} +
    \left( 1-\bm1^\intercal\bm\alpha \right)
    (\bar{\bm p}^d)^\intercal\eta
    \notag\\
    &= \bar p_0^{\mathrm{sl}} + \alpha_0(\bar{\bm p}^d)^\intercal\eta.
    \label{eq:feeder-head-active-response}
\end{align}
Thus, $\alpha_0$ parameterizes the fraction of the aggregate disturbance supplied through feeder-head exchange, while $p_0^{\mathrm{sl}}$ remains an algebraic network quantity.

The DER participation law in \eqref{eq:active-policy} acts directly on the disturbance and is therefore a disturbance-feedforward policy, in contrast to the local Volt/VAR output feedback of Section~\ref{subsec:outputfeedback}. The feeder-head exchange is the
residual balancing response implied by the feeder power balance.

\begin{figure}[htbp]
    \centering
    \includegraphics[scale=1]{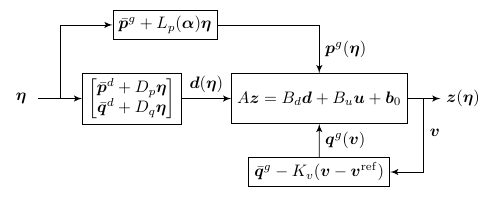}
    \caption{Control architecture for the radial distribution feeder. The disturbance $\bm{\eta}$ drives the load model and feedforward active-power participation policy, while reactive-power control is generated through local voltage feedback. The feeder-head active-power exchange $p_0^{\mathrm{sl}}$ is a coordinate of $\bm z(\bm\eta)$ and captures the residual active-power imbalance supplied by the upstream network.}
    \label{fig:control-architecture}
\end{figure}

Substituting the load model and the active-power policy into
\eqref{eq:after-fb} yields 
\begin{gather*}
    A_{\mathrm{cl}}(K_v) \bm z = B_d \begin{bmatrix}
        \bar{\bm p}^d + D_p \bm \eta \\ 
        \bar{\bm q}^d + D_q \bm \eta
    \end{bmatrix} - B_p (\bar{\bm p}^g + L_p(\bm \alpha) \bm \eta) + \bm b_{\mathrm{cl}}(K_v).
\end{gather*}
Separating the nominal and disturbance-dependent terms yields
\begin{gather}
    A_{\mathrm{cl}}(K_v) \bm z = \bar{\bm b}_{\mathrm{cl}}(K_v)  + G_{\mathrm{cl}}(\bm \alpha) \bm \eta, \label{eq:full}
\end{gather}
where, 
\begin{gather*}
    \bar{\bm b}_{\mathrm{cl}}(K_v) \triangleq B_d \begin{bmatrix}
        \bar{\bm p}^d  \\ 
        \bar{\bm q}^d
    \end{bmatrix} - B_p \bar{\bm p}^g + \bm b_{\mathrm{cl}}(K_v), \\ 
    G_{\mathrm{cl}}(\bm \alpha) \triangleq B_d \begin{bmatrix}
        D_p \\ D_q 
    \end{bmatrix} - B_p L_p(\bm \alpha)
\end{gather*}
The model in \eqref{eq:full} separates the two control mechanisms. The Volt/VAR gains $K_v$ modify the algebraic operator through local output feedback, whereas $\bm\alpha$ modifies the disturbance map through active-power feedforward. The residual share $\alpha_0=(1-\bm1^\intercal\bm\alpha)$ is supplied through feeder-head exchange. Although $\alpha_0$ does not appear explicitly in $G_{\mathrm{cl}}$, its effect is contained in the $p_0^{\mathrm{sl}}$ component of the resulting state response. Figure~~\ref{fig:control-architecture} summarizes the resulting closed-loop control architecture.

The non-singularity of $A^{\mathrm{cl}}(K_v)$, established in Proposition \ref{prop:Acl} in the Appendix, allows the closed-loop feeder model in \eqref{eq:full} to be written explicitly as an affine function of the disturbance $\bm \eta$. In particular 
\begin{gather}
    \bm z = \underbrace{\left(A^{\mathrm{cl}}(K_v)\right)^{-1} \bar{\bm b}_{\mathrm{cl}}(K_v)}_{\triangleq~ \bar {\bm z}_{\mathrm{cl}}(K_v)} + \underbrace{\left(A^{\mathrm{cl}}(K_v)\right)^{-1} G_{\mathrm{cl}}(\alpha)}_{\triangleq~ \Psi_{\mathrm{cl}}(K_v, \bm \alpha)} \bm \eta. \label{eq:affine-full}
\end{gather}
Equivalently,
\begin{gather}
    \bm z (\bm \eta) = \bar {\bm z}_{\mathrm{cl}}(K_v) + \Psi_{\mathrm{cl}}(K_v, \bm \alpha) \bm \eta. \label{eq:affine}
\end{gather}
Since the balancing action vanishes at $\bm \eta=0$, the nominal closed-loop state $\bar{\bm z}_{\mathrm{cl}}(K_v)$ is independent of $(\alpha_0,\bm\alpha)$; the participation factors affect only the disturbance response. In particular, the row of $\Psi_{\mathrm{cl}}(K_v,\bm\alpha)$ corresponding to $p_0^{\mathrm{sl}}$ is $\alpha_0(\bar{\bm p}^d)^\intercal$. Thus, $\Psi_{\mathrm{cl}}$ captures the disturbance sensitivities of branch flows, bus voltages, and feeder-head active- and reactive-power exchanges. Thus, every linear constraint on the algebraic state can be written as an affine function of $\bm \eta$, enabling the robust formulations developed in Section~\ref{sec:robust}.

\subsection{Network and actuator limits} \label{subsec:limits}
The closed-loop feeder in \eqref{eq:affine} must satisfy prescribed network and actuator limits for every disturbance
$\bm\eta\in\mathcal U(\rho)$. These include voltage, controllable active- and reactive-power injection, feeder-head interchange and balancing, and branch-flow limits. Unless stated otherwise, all disturbance-dependent inequalities below are required to hold for every $\bm\eta\in\mathcal U(\rho)$.

For each non-root bus $j\in\mathcal N$, the squared voltage magnitude must satisfy
\begin{gather}
    v_j^{\min} \leqslant v_j(\bm \eta) \leqslant v_j^{\max} \quad \forall j \in \mathcal N, ~~\bm \eta \in \mathcal U(\rho). \label{eq:v-limits}
\end{gather}
 
Let $\mathcal{N}_G\subseteq\mathcal{N}$ denote the set of buses equipped with controllable active-power resources. Their injections are constrained by
\begin{gather}
    p^{g,\min}_j \leqslant p_j^g(\bm \eta) \leqslant p^{g,\max}_j \quad \forall j \in \mathcal N_G, ~~\bm \eta \in \mathcal U(\rho). \label{eq:pg-limits}
\end{gather}
Similarly, for each bus $j\in\mathcal N_I$ governed by the local Volt/VAR law, the reactive-power injection must satisfy
\begin{gather}
    q^{g,\min}_j \leqslant q_j^g(\bm \eta) \leqslant q^{g,\max}_j \quad \forall j \in \mathcal N_I, ~~\bm \eta \in \mathcal U(\rho). \label{eq:qg-limits}
\end{gather}
Thus, the affine Volt/VAR law is certified only over disturbances for which the prescribed reactive-power capability limits are respected. At buses without voltage-responsive reactive-power control, fixed reactive-power set points are assumed to satisfy their corresponding capability limits.

We also constrain the active- and reactive-power exchanges with the upstream network. Since $p_0^{\mathrm{sl}}$ and
$q_0^{\mathrm{sl}}$ are retained explicitly in the algebraic state,
\begin{subequations}
    \begin{gather}
    p_0^{\mathrm{sl},\min} \leqslant
    p_0^{\mathrm{sl}}(\bm\eta) \leqslant
    p_0^{\mathrm{sl},\max},
    \label{eq:feeder-head-p-limit}\\
    q_0^{\mathrm{sl},\min} \leqslant
    q_0^{\mathrm{sl}}(\bm\eta) \leqslant
    q_0^{\mathrm{sl},\max}.
    \label{eq:feeder-head-q-limit}
    \end{gather}
    \label{eq:feeder-head-interchange}
\end{subequations}
These bounds represent absolute limits on feeder-head interchange.

We also distinguish the upstream network's available active-power balancing authority from its absolute interchange limits. Let $r_0^{\uparrow}\geqslant0$ and $r_0^{\downarrow}\geqslant0$ denote the maximum permitted increase and decrease, respectively, in feeder-head active-power import relative to the nominal exchange $\bar p_0^{\mathrm{sl}}$. We impose
\begin{gather}
    -r_0^{\downarrow} \leqslant p_0^{\mathrm{sl}}(\bm\eta)-\bar p_0^{\mathrm{sl}} \leqslant r_0^{\uparrow}.
    \label{eq:feeder-head-balancing}
\end{gather}
Hence, \eqref{eq:feeder-head-interchange} limits the absolute feeder-head operating point, whereas \eqref{eq:feeder-head-balancing} limits the incremental support available from the upstream network. For the one-sided demand-growth set $\mathcal U_+(\rho)$, \eqref{eq:feeder-head-active-response} implies
\begin{gather*}
    p_0^{\mathrm{sl}}(\bm\eta)-\bar p_0^{\mathrm{sl}}
    = \alpha_0(\bar{\bm p}^d)^\intercal\bm\eta \geqslant 0,
\end{gather*}
so the downward balancing limit is non-binding; both directions are relevant for the symmetric disturbance set $\mathcal U_2(\rho)$.

Finally, the branch power flows must remain within their prescribed limits. Since $p_j(\bm{\eta})$ and $q_j(\bm{\eta})$ denote the active- and reactive-power flows on branch $(\pi(j),j)\in\mathcal{E}$, we impose
\begin{subequations}
\begin{gather}
    -t_j^{\max} \leqslant p_j(\bm \eta) \leqslant t_j^{\max} \quad \forall j \in \mathcal N, ~~\bm \eta \in \mathcal U(\rho), \label{eq:p-limits} \\ 
    -t_j^{\max} \leqslant q_j(\bm \eta) \leqslant t_j^{\max} \quad \forall j \in \mathcal N, ~~\bm \eta \in \mathcal U(\rho). \label{eq:q-limits}
\end{gather}
    \label{eq:thermal-limits}
\end{subequations}
We remark that rather than imposing the quadratic branch thermal limit 
\begin{gather*}
    p_j(\bm{\eta})^2+q_j(\bm{\eta})^2\leqslant  (t_j^{\max})^2,
\end{gather*}
we impose separate linear bounds on the active- and reactive-power flows to preserve the linear structure of the network constraints. This choice leads to a polyhedral approximation of the quadratic thermal-limit region and simplifies the subsequent robustness analysis. If a more accurate representation of the thermal limit is desired, additional linear inequalities may be introduced to obtain an increasingly tight polyhedral approximation of the quadratic constraint.

Collectively, \eqref{eq:v-limits}--\eqref{eq:thermal-limits} define the closed-loop feasible disturbance region used in the flexibility-certification problem of Section~\ref{sec:robust}.

\section{Robust Flexibility Certification} \label{sec:robust}
The affine closed-loop model developed in Section \ref{sec:modeling} maps the load disturbance $\bm\eta$ to the feeder state under the Volt/VAR feedback and active-power feedforward control architecture. We use this representation to formulate a robust flexibility-certification problem; for the control architecture in Figure~ \ref{fig:control-architecture}, what is the largest uncertainty-set radius $\rho$ for which all network and actuator limits remain satisfied for every $\bm\eta\in\mathcal U(\rho)$? Under the one-sided uncertainty set $\mathcal U(\rho)\equiv\mathcal U_+(\rho)$ adopted here, this radius quantifies robust demand-growth capability. We first derive the flexibility margin for fixed $(K_v,\bm\alpha)$ and then show that, with $K_v$ fixed, the active-power participation factors $\bm\alpha$ can be optimized jointly with $\rho$ through an exact SOC reformulation. Joint optimization of the Volt/VAR gains $K_v$ and the flexibility margin is nonconvex because $K_v$ enters the closed-loop disturbance response bilinearly; we therefore treat $K_v$ as prescribed and leave joint Volt/VAR controller synthesis to future work.

\subsection{Flexibility margin for fixed controller} \label{subsec:rho-opt}
For fixed $(K_v,\bm{\alpha})$, with the feeder-head participation $\alpha_0 = 1- \bm 1^\intercal \bm \alpha \geqslant 0$ determined by $\bm \alpha$, both $\bar{\bm z}_{\mathrm{cl}}(K_v)$ and the disturbance-response matrix $\Psi_{\mathrm{cl}}(K_v,\bm{\alpha})$ are fixed. We assume throughout that the nominal closed-loop operating point is feasible. Consequently, every scalar quantity appearing in the engineering constraints \eqref{eq:v-limits}--\eqref{eq:thermal-limits} can be expressed in the generic form 
\begin{gather}
    y_i(\bm \eta) = \bar{y}_i + \bm \psi_i^\intercal \bm \eta, \quad \forall i \in \mathcal I \label{eq:generic-state}
\end{gather}
where $\mathcal I$ indexes all scalar network and actuator limits. This includes the voltage, generation, branch-flow, absolute feeder-head
interchange, and incremental feeder-head balancing limits. For the incremental feeder-head balancing constraint,
for example, $y_i(\bm\eta)=p_0^{\mathrm{sl}}(\bm\eta)-\bar
p_0^{\mathrm{sl}}$, with $y_i^{\min}=-r_0^\downarrow$ and
$y_i^{\max}=r_0^\uparrow$.

For quantities contained directly in the algebraic state $\bm z(\bm \eta)$ in \eqref{eq:affine}, $\bar{y}_i$ is the corresponding component of $\bar{\bm z}_{\mathrm{cl}}$, and $\bm \psi_i^\intercal$ is the corresponding row of $\Psi_{\mathrm{cl}}$. The active- and reactive-power generation quantities, although not components of $\bm z(\bm\eta)$, also admit the affine representation in \eqref{eq:generic-state}. In particular, from the active-power participation policy,
\begin{gather*}
p_j^g(\bm\eta) = \bar p_j^g + \alpha_j(\bar{\bm p}^{d})^\intercal\bm\eta ~~\Rightarrow~~
\bar y_{p_j^g}=\bar p_j^g, \quad \bm\psi_{p_j^g} = \alpha_j\bar{\bm p}^{d}.
\end{gather*}
Similarly, let $\bar v_j$ denote the $j$th voltage component of $\bar{\bm z}_{\mathrm{cl}}$ and let $\bm\psi_{v,j}^\intercal$ denote the corresponding row of $\Psi_{\mathrm{cl}}$. Then the Volt/VAR output feedback policy in Section \ref{subsec:outputfeedback} yields
\begin{align*}
q_j^g(\bm\eta) &= \bar q_j^g - k_j (v_j(\bm\eta)-v_j^{\mathrm{ref}} ) \\
& = \bar q_j^g - k_j(\bar v_j-v_j^{\mathrm{ref}}) - k_j\bm\psi_{v,j}^\intercal\bm\eta,
\end{align*}
and hence
\begin{gather*}
\bar y_{q_j^g} = \bar q_j^g-k_j(\bar v_j-v_j^{\mathrm{ref}}), \qquad
\bm\psi_{q_j^g} = -k_j\bm\psi_{v,j}.
\end{gather*}

For fixed controller parameters, define the closed-loop feasible
disturbance region as
\begin{gather}
    \mathcal F(K_v, \bm \alpha) \triangleq \{ \bm \eta: y_i^{\min} \leqslant y_i(\bm \eta) \leqslant y_i^{\max}, ~~\forall i \in \mathcal I\}. \label{eq:flex-set}
\end{gather}
The flexibility margin is the largest scaling of the prescribed uncertainty set contained in this region,
\begin{gather}
    \rho^\star(K_v, \bm \alpha) \triangleq \operatorname{sup} \{ \rho \geqslant 0: \mathcal U(\rho) \subseteq \mathcal F(K_v, \bm \alpha) \}. \label{eq:rho-fixed-controller}
\end{gather}
For $\mathcal U(\rho)\equiv\mathcal U_+(\rho)$, this quantity is the robust demand-growth flexibility margin; every disturbance $\bm\eta\in\mathcal U(\rho^\star)$ is certified to satisfy all network and actuator limits under the prescribed controller and the adopted LinDistFlow model.

To characterize $\rho^\star$, define componentwise 
\begin{gather*}
    [\bm \psi]_+ \triangleq \max(\bm \psi, \bm 0), \qquad [\bm \psi]_- \triangleq \max(-\bm \psi,\bm 0).
\end{gather*}
Consider a generic two-sided engineering limit
\begin{gather*}
     y_i^{\min} \leqslant \bar{y}_i + \bm \psi_i^\intercal \bm \eta \leqslant y_i^{\max}, \quad \bm \eta \in \mathcal U(\rho).
\end{gather*}
For the one-sided Euclidean set in \eqref{eq:eta-ball},
\begin{gather*}
    \max_{\bm \eta \in \mathcal U(\rho)} ~\bm \psi_i^\intercal \bm \eta = \rho \| [\bm \psi_i]_+ \|_2, \quad 
    \min_{\bm \eta \in \mathcal U(\rho)} ~\bm \psi_i^\intercal \bm \eta = -\rho \| [\bm \psi_i]_- \|_2
\end{gather*}
These expressions follow from the support function of the Euclidean
ball restricted to the non-negative orthant. Hence, the semi-infinite
engineering constraint has the exact robust counterpart
\begin{gather*}
    \bar{y}_i + \rho \|[\bm \psi_i]_+ \|_2 \leqslant y_i^{\max} \text{ and } \bar y_i - \rho \|[\bm \psi_i]_-\|_2 \geqslant  y_i^{\min}.
\end{gather*}
Therefore, for fixed $(K_v,\bm\alpha)$, the flexibility margin is obtained from
\begin{subequations}
\begin{flalign}
\rho^\star(K_v,\bm{\alpha})
= & \max_{\rho \geqslant 0}\quad
 \rho ~~~ \text{s.t.}&& \nonumber\\
\quad
& \bar y_i + \rho \|[\bm{\psi}_i]_+\|_2 \leqslant y_i^{\max}, \quad \forall i\in\mathcal I, && \label{eq:robust-upper}\\
& \bar y_i - \rho \|[\bm{\psi}_i]_-\|_2 \geqslant y_i^{\min}, \quad \forall i\in\mathcal I. && \label{eq:robust-lower}
\end{flalign}
\label{eq:fixed-controller-opt}
\end{subequations}
Since $\bar y_i$ and $\bm\psi_i$ are fixed, \eqref{eq:fixed-controller-opt} is a linear program in the single variable $\rho$. Equivalently,
\begin{gather*}
    \rho^\star(K_v,\bm{\alpha}) = \min_{i \in \mathcal I} \left\{ \frac{y_i^{\max} - \bar y_i}{\|[\bm{\psi}_i]_+\|_2}, \frac{\bar y_i - y_i^{\min}}{\|[\bm{\psi}_i]_-\|_2} \right\}.
\end{gather*}
where the term with zero denominator is interpreted as $+\infty$; such a limit is non-binding because nominal feasibility has been assumed.
For the symmetric uncertainty set $\mathcal U_2(\rho)$ in \eqref{eq:eta-ball-euc}, the same development applies with $\|[\bm\psi_i]_+\|_2$ and $\|[\bm\psi_i]_-\|_2$ both replaced by $\|\bm\psi_i\|_2$.

\subsection{Optimization of active-power participation factors} \label{subsec:alpha-rho-opt}
The fixed-controller formulation in \eqref{eq:fixed-controller-opt} quantifies the flexibility margin for prescribed active-power participation factors. We now optimize the downstream participation vector $\bm\alpha$ and the feeder-head participation $\alpha_0$ jointly with the flexibility margin. From \eqref{eq:alpha-policy},
\begin{gather*}
    \alpha_0 \geqslant 0,\qquad
    \bm\alpha\geqslant\bm 0,\qquad
    \alpha_0+\bm 1^\intercal\bm\alpha=1,
\end{gather*}
with $\alpha_j=0$ at buses without participating active-power resources. Equivalently, $\alpha_0=1-\bm 1^\intercal\bm\alpha$ is determined by the downstream participation factors.

Unlike the Volt/VAR gains $K_v$, the active-power participation factors do not modify the closed-loop algebraic operator $A^{\mathrm{cl}}(K_v)$; rather, they enter through the disturbance feedforward map $G_{\mathrm{cl}}(\bm{\alpha})$. Define 
\begin{gather*}
    G_0 \triangleq B_d \begin{bmatrix}
        D_p \\ D_q 
    \end{bmatrix}.
\end{gather*}
Then 
    $G_{\mathrm{cl}}(\bm \alpha) = G_0 - B_p \bm \alpha (\bar{\bm p}^d)^\intercal$,
and for fixed $K_v$, the disturbance-response matrix satisfies
\begin{gather}
    A^{\mathrm{cl}}(K_v) \Psi_{\mathrm{cl}}(K_v, \bm \alpha) = G_0 - B_p \bm \alpha (\bar{\bm p}^d)^\intercal. \label{eq:disturbance-response}
\end{gather}
Hence, $\Psi_{\mathrm{cl}}(K_v,\bm{\alpha})$ is affine in $\bm{\alpha}$.

Joint optimization of $\rho$ and $\bm\alpha$ in \eqref{eq:fixed-controller-opt} introduces products of the form
\begin{gather*}
    \rho \|[\bm \psi_i (\bm \alpha)]_+\|_2, \qquad \rho \|[\bm \psi_i (\bm \alpha)]_-\|_2.
\end{gather*}
We eliminate this coupling through scaled variables 
\begin{gather*}
    \bm\beta \triangleq \rho\bm\alpha,
    \qquad
    \beta_0 \triangleq \rho\alpha_0,
    \qquad
    \hat\Psi_{\mathrm{cl}} \triangleq \rho\Psi_{\mathrm{cl}}(K_v,\bm\alpha).
\end{gather*}
The participation constraints in \eqref{eq:alpha-policy} become
\begin{gather}
    \beta_0\geqslant0,~
    \bm\beta\geqslant\bm 0,~
    \beta_0+\bm 1^\intercal\bm\beta=\rho,~
    \beta_j=0, \forall j\notin\mathcal N_G .
    \label{eq:beta-control-law}
\end{gather}
Moreover, multiplying \eqref{eq:disturbance-response} by $\rho$ yields
\begin{gather}
    A^{\mathrm{cl}}(K_v) \hat\Psi_{\mathrm{cl}} = \rho G_0 - B_p \bm \beta (\bar{\bm p}^d)^\intercal. \label{eq:disturbance-response-transformed}
\end{gather}
For fixed $K_v$, \eqref{eq:disturbance-response-transformed} is affine
in $\rho$, $\bm\beta$, and $\hat\Psi_{\mathrm{cl}}$. For each constrained quantity
\begin{gather*}
     y_i(\bm \eta) =  \bar{y}_i + \left(\bm \psi_i(\bm \alpha)\right)^\intercal \bm \eta,
\end{gather*}
define the corresponding scaled disturbance sensitivity as 
\begin{gather}
    \hat{\bm \psi}_i \triangleq \rho \bm \psi_i(\bm \alpha). \label{eq:scaled-psi}
\end{gather} 
For quantities contained in the algebraic state, $\hat{\bm{\psi}}_i^\intercal$ is the corresponding row of $\hat{\Psi}_{\mathrm{cl}}$. For the active-power injection at generator bus $j$,
\begin{gather*}
    \hat{\bm\psi}_{p_j^g} = \beta_j\bar{\bm p}^{d},
\end{gather*}
while the Volt/VAR law gives
\begin{gather*}
    \hat{\bm\psi}_{q_j^g} = -k_j\hat{\bm\psi}_{v,j}.
\end{gather*}
The scaled sensitivity of the incremental feeder-head active-power exchange is
\begin{gather*}
    \hat{\bm\psi}_{p_0^{\mathrm{sl}}-\bar p_0^{\mathrm{sl}}} = \beta_0\bar{\bm p}^{d},
\end{gather*}
consistent with $\beta_0=\rho-\bm 1^\intercal\bm\beta$. Thus, every scaled disturbance sensitivity is affine in the reformulated decision variables. 

Using \eqref{eq:scaled-psi}, the exact robust counterpart of the constraints in \eqref{eq:fixed-controller-opt} can be expressed for every $i \in \mathcal I$ as 
\begin{gather}
    \bar y_i + \| [\hat{\bm \psi}_i]_+\|_2 \leqslant y_i^{\max} \text{ and } \bar y_i - \|[\hat {\bm \psi}_i]_-\|_2 \geqslant y_{i}^{\min}. \label{eq:scaled-limits}
\end{gather}
Introduce auxiliary variables $\bm s_i^+,\bm s_i^-\in\mathbb R_+^n$. The constraints in
\eqref{eq:scaled-limits} are equivalently represented for every $i\in\mathcal I$ by
\begin{gather}
    \begin{gathered}
        \bm s_i^+ \geqslant \hat{\bm \psi}_i,\quad \bm s_i^+ \geqslant \bm 0,\quad~~
\|\bm s_i^+\|_2 \leqslant y_i^{\max}-\bar y_i, \\ 
        \bm s_i^- \geqslant -\hat{\bm \psi}_i,\quad \bm s_i^- \geqslant \bm 0,\quad
\|\bm s_i^-\|_2 \leqslant \bar y_i-y_i^{\min}.
    \end{gathered}
    \label{eq:scaled-limits-soc}
\end{gather}
Here, each $\hat{\bm\psi}_i$ is the corresponding affine function of $\hat{\Psi}_{\mathrm{cl}}$, $\bm\beta$, and $\beta_0$ described above. Combining \eqref{eq:beta-control-law}, \eqref{eq:disturbance-response-transformed}, and
\eqref{eq:scaled-limits-soc} yields the following optimization problem:
\begin{subequations}
\begin{flalign}
\rho^\star(K_v) = & \max_{\rho,\beta_0, \bm \beta, \hat{\Psi}_\mathrm{cl},\bm s_i^\pm} \quad \rho ~~ \text{s.t.} && \label{eq:participation-soc-obj} \\
\quad & A^{\mathrm{cl}}(K_v) \hat{\Psi}_{\mathrm{cl}} = \rho G_0 - B_p\bm{\beta}(\bar{\bm p} ^d)^\intercal, && \label{eq:participation-soc-physics} \\
& \bar y_i + \| \bm s_i^+ \|_2 \leqslant y_i^{\max}, \qquad \forall i\in\mathcal I, && \label{eq:participation-soc-upper} \\
& \bar y_i - \| \bm s_i^- \|_2 \geqslant y_i^{\min}, \qquad \forall i\in\mathcal I, && \label{eq:participation-soc-lower} \\
& \bm s_i^{+} \geqslant \bm 0, ~\bm s_i^- \geqslant \bm 0, \qquad ~~~\forall i \in \mathcal I, && \label{eq:participation-soc-lb} \\ 
& \bm s_i^+ \geqslant \hat{\bm \psi}_i, ~\bm s_i^- \geqslant -\hat{\bm \psi}_i, ~~~\forall i \in \mathcal I, && \label{eq:participation-soc-spsi} \\
& \beta_0\geqslant0,~
\bm\beta\geqslant\bm 0,~
\beta_0+\bm 1^\intercal\bm\beta=\rho, && \label{eq:participation-soc-simplex} \\
& \beta_j=0,~ \forall j\notin\mathcal N_G,
\quad \rho\geqslant0. &&
\label{eq:participation-soc-nongen}
\end{flalign}
\label{eq:vary-alpha-opt}
\end{subequations}
\begin{proposition}[Exact SOCP reformulation] \label{prop:soc}
For fixed $K_v$, \eqref{eq:vary-alpha-opt} is an exact SOCP reformulation of the joint optimization of the flexibility margin and the active-power participation factors satisfying \eqref{eq:alpha-policy}. For $\rho^\star(K_v)>0$, the optimal participation factors are recovered as
\begin{gather*}
    \bm\alpha^\star = \frac{\bm\beta^\star}{\rho^\star(K_v)}, \qquad
    \alpha_0^\star = \frac{\beta_0^\star}{\rho^\star(K_v)}.
\end{gather*}
\end{proposition}
\begin{proof}
See Appendix~\ref{subsec:soc-proof}.
\end{proof}

\begin{remark}[Polyhedral uncertainty sets] The SOCP structure in Proposition~\ref{prop:soc} results from the Euclidean geometry of the uncertainty set in~\eqref{eq:eta-ball}. In particular, consider instead the one-sided infinity-norm uncertainty set \begin{equation} 
    \mathcal{U}_{\infty}^{+}(\rho) \triangleq \left\{ \eta \in \mathbb{R}^{n} : \|\bm \eta\|_{\infty} \leqslant \rho,\; \bm \eta \geqslant  \bm 0 \right\}. 
\end{equation} 
For any sensitivity vector $\bm \psi_i$, its support function over $\mathcal{U}_{\infty}^{+}(\rho)$ is 
\begin{equation} 
    \max_{\bm \eta \in \mathcal{U}_{\infty}^{+}(\rho)} \bm \psi_i^{\intercal} \bm \eta = \rho \|[\bm \psi_i]_{+}\|_{1}, 
\end{equation} 
while 
\begin{equation} 
    \min_{\bm \eta \in \mathcal{U}_{\infty}^{+}(\rho)} \bm \psi_i^{\intercal}\bm \eta = -\rho \|[\bm \psi_i]_{-}\|_{1}. 
\end{equation} 
Hence, after introducing the scaled sensitivity $\widehat{\bm \psi}_i = \rho \bm \psi_i$, the robust counterparts become \begin{equation} 
    \bar{y}_i + \|[\widehat{\bm \psi}_i]_{+}\|_{1} \leqslant y_i^{\max}, \qquad \bar{y}_i - \|[\widehat{\bm \psi}_i]_{-}\|_{1} \geqslant y_i^{\min}. 
\end{equation} 
Since the $\ell_1$-norm admits an exact linear epigraph representation, these constraints can be represented using only linear inequalities. Consequently, for fixed $K_v$, the joint optimization of the flexibility margin and the active-power participation factors becomes a linear program rather than an SOCP. The same observation applies to the structured feedforward formulation in the subsequent. Thus, the optimization class is determined in part by the geometry chosen for the prescribed uncertainty set: the Euclidean set leads to SOC constraints, whereas the infinity-norm set leads to linear constraints. 
\end{remark}

\begin{remark}[Joint optimization of Volt/VAR gains $K_v$ and $\rho$] 
If $K_v$ is also optimized, $A^{\mathrm{cl}}(K_v)\hat{\Psi}_{\mathrm{cl}}$ and the nominal closed-loop equations contain bilinear terms. Joint Volt/VAR gain design therefore leads to a non-convex static output-feedback synthesis problem \cite{syrmos1994bilinear}.
\end{remark}

\section{Structured Feedforward Control} \label{sec:comm}
The global participation policy in \eqref{eq:active-policy} requires each participating downstream generator to respond to the same aggregate load disturbance, with the feeder head supplying the residual imbalance. In practice, DER coordination may be limited by measurement and communication availability \cite{magnusson2019voltage,liu2018hybrid}, so a generator may have access only to disturbances at a subset of load buses. We therefore generalize the rank-one feedforward map $L_p(\bm\alpha)$ to a structured load-to-generation map whose sparsity pattern encodes the available disturbance information. This preserves the affine closed-loop disturbance response while allowing communication-constrained feedforward control.

\subsection{Load-to-generator participation policy} \label{subsec:policy}
Let $\mathcal{N}_L \subseteq \mathcal{N}$ denote the set of buses with non-zero nominal active-power demand, and recall that $\mathcal{N}_G \subseteq \mathcal{N}$ denotes the set of buses with controllable active-power generation. We generalize the participation vector $\bm{\alpha}$ to a participation matrix
\begin{gather*}
    \Gamma \triangleq [\gamma_{ij}] \in \mathbb R^{n \times n}, 
\end{gather*}
where $\gamma_{ij}$ denotes the fraction of the active-power disturbance at load bus $j$ assigned to the controllable generator at bus $i$. Similarly, let $\gamma_{0j}$ denote the fraction supplied through the feeder head, and define
\begin{gather*}
    \bm\gamma_0 \triangleq \operatorname{col}(\gamma_{0j})_{j\in\mathcal N}.
\end{gather*}
The participation factors satisfy
\begin{gather}
    \gamma_{0j}\geqslant 0,\quad 
    \gamma_{ij}\geqslant 0,\quad
    \gamma_{0j} + \sum_{i\in\mathcal N_G}\gamma_{ij} = 1,~ \forall j\in\mathcal N_L,
    \label{eq:gamma-policy}
\end{gather}
with $\gamma_{ij}=0$ whenever $i\notin\mathcal N_G$ or $j\notin\mathcal N_L$, and $\gamma_{0j}=0$ for $j\notin\mathcal N_L$.
The resulting downstream active-power feedforward policy is
\begin{gather}
    \bm p^g(\bm \eta) = \bm {\bar p}^g + \Gamma D_p \bm \eta. \label{eq:gamma-control-law}
\end{gather}
Thus, the $j$th column of $\Gamma$ specifies how the active-power disturbance at load bus $j$ is distributed among the controllable generators, while $\gamma_{0j}$ specifies the residual share supplied by the upstream network. From \eqref{eq:gamma-policy}, the corresponding feeder-head exchange is
\begin{gather*}
    p_0^{\mathrm{sl}}(\bm\eta) = \bar p_0^{\mathrm{sl}} + \bm\gamma_0^\intercal D_p\bm\eta.
\end{gather*}
The global participation policy in \eqref{eq:active-policy} is recovered as the special case
\begin{gather*}
    \Gamma = \bm\alpha\bm 1^\intercal,
    \qquad
    \bm\gamma_0 = \alpha_0\bm 1,
\end{gather*}
for which $\Gamma D_p=L_p(\bm\alpha)$ and every load disturbance is balanced using the same participation factors $(\alpha_0,\bm\alpha)$. The structured policy therefore allows the balancing response to depend on the location of the disturbance.

\subsection{Communication-constrained controller class} \label{subsec:controller}
Let $M\triangleq[m_{ij}]\in\{0,1\}^{n\times n}$ denote a prescribed communication mask, where $m_{ij}=1$ if the controllable generator at bus $i$ has access to the disturbance at load bus $j$. The admissible participation matrices satisfy
\begin{gather}
    \gamma_{ij}=0
    \qquad \text{whenever } m_{ij}=0.
    \label{eq:communication-mask}
\end{gather}
Thus, the sparsity pattern of $M$ specifies the disturbance information available to each downstream controller, while the non-zero entries of $\Gamma$ determine the corresponding balancing response. The feeder-head participation vector $\bm\gamma_0$ represents the residual balancing supplied by the upstream network and is not restricted by $M$.

A useful choice is an $r$-hop communication mask $M^{(r)}$, defined by
\begin{gather}
    m_{ij}^{(r)} \triangleq
    \begin{cases}
        1, & i\in\mathcal N_G,\ j\in\mathcal N_L,\ 
        \operatorname{dist}(i,j)\leqslant r,\\
        0, & \text{otherwise},
    \end{cases}
    \label{eq:r-hop-mask}
\end{gather}
where $\operatorname{dist}(i,j)$ is the graph distance between buses $i$ and $j$ on the feeder. Increasing $r$ enlarges the admissible controller class, while $M^{(\infty)}$ corresponds to full load-to-generator information.

\subsection{Communication-constrained flexibility} \label{subsec:reformulation}
We now optimize the structured feedforward policy subject to a prescribed communication mask $M$. Replacing the global map $L_p(\bm\alpha)$ by $\Gamma D_p$ results in the closed-loop disturbance-response equation
\begin{gather}
    A^{\mathrm{cl}}(K_v) \Psi_{\mathrm{cl}}(K_v,\Gamma) = G_0-B_p\Gamma D_p.
    \label{eq:disturbance-response-gamma}
\end{gather}
Thus, for fixed $K_v$, the disturbance response remains affine in the feedforward policy. Following the scaled-variable construction in
Section \ref{subsec:alpha-rho-opt}, define
\begin{gather*}
    \Theta\triangleq\rho\Gamma,\qquad
    \bm\theta_0\triangleq\rho\bm\gamma_0,\qquad
    \hat\Psi_{\mathrm{cl}} \triangleq \rho\Psi_{\mathrm{cl}}(K_v,\Gamma),
\end{gather*}
where $\Theta=[\theta_{ij}]$ and $\bm\theta_0=\operatorname{col}(\theta_{0j})_{j\in\mathcal N}$. Then \eqref{eq:disturbance-response-gamma} becomes
\begin{gather}
    A^{\mathrm{cl}}(K_v)\hat\Psi_{\mathrm{cl}} = \rho G_0-B_p\Theta D_p.
    \label{eq:disturbance-response-gamma-transformed}
\end{gather}
The participation constraints in \eqref{eq:gamma-policy} scale to
\begin{gather*}
    \theta_{0j}\geqslant0,\quad
    \theta_{ij}\geqslant0,\quad
    \theta_{0j}+\sum_{i\in\mathcal N_G}\theta_{ij}= \rho,
    ~ \forall j\in\mathcal N_L,
\end{gather*}
while the information constraint requires $\theta_{ij}=0$ whenever $m_{ij}=0$.

Let $\hat{\bm\psi}_i$ denote the scaled disturbance sensitivity defined
as in \eqref{eq:scaled-psi}. For quantities contained in the algebraic
state, $\hat{\bm\psi}_i^\intercal$ is the corresponding row of
$\hat\Psi_{\mathrm{cl}}$; the active-power generation sensitivities are
given by the rows of $\Theta D_p$. The scaled sensitivity of the
incremental feeder-head exchange is
\begin{gather*}
    \hat{\bm\psi}_{p_0^{\mathrm{sl}}-\bar p_0^{\mathrm{sl}}}
    =
    D_p\bm\theta_0.
\end{gather*}
Hence, all scaled sensitivities are affine functions of the reformulated
decision variables.

Using the SOC representation in \eqref{eq:scaled-limits-soc}, the
communication-constrained flexibility margin is obtained from
\begin{subequations}
\begin{flalign}
\rho^\star(K_v,M)
= {}& \max_{\rho,\Theta,\bm\theta_0,\hat{\Psi}_{\mathrm{cl}}, \bm s_i^\pm}
\quad \rho \quad \text{s.t.} && \label{eq:comm-soc-obj}\\
& A^{\mathrm{cl}}(K_v)\hat{\Psi}_{\mathrm{cl}} = \rho G_0-B_p\Theta D_p, && \label{eq:comm-soc-physics}\\
& \bar y_i+\|\bm s_i^+\|_2 \leqslant y_i^{\max}, \quad \forall i\in\mathcal I,
&& \label{eq:comm-soc-upper}\\
& \bar y_i-\|\bm s_i^-\|_2 \geqslant y_i^{\min}, \quad \forall i\in\mathcal I,
&& \label{eq:comm-soc-lower}\\
& \bm s_i^+\geqslant\bm 0,\quad \bm s_i^-\geqslant\bm 0, \quad \forall i\in\mathcal I,
&& \label{eq:comm-soc-lb}\\
& \bm s_i^+\geqslant\hat{\bm\psi}_i,\quad \bm s_i^-\geqslant-\hat{\bm\psi}_i,
\quad \forall i\in\mathcal I, 
&& \label{eq:comm-soc-spsi}\\
& \theta_{0j} + \sum_{i\in\mathcal N_G}\theta_{ij} = \rho,
\quad \forall j\in\mathcal N_L,
&&
\label{eq:comm-soc-participation}\\
& \theta_{ij}=0, \quad \forall(i,j):m_{ij}=0,
&& \label{eq:comm-soc-sparsity}\\
& \theta_{0j}\geqslant0,~ \forall j\in\mathcal N_L,~~
\theta_{ij}\geqslant0,~~ \rho\geqslant0.
&&
\label{eq:comm-bounds}
\end{flalign}
\label{eq:comm-opt}
\end{subequations}
For fixed $K_v$ and $M$, the participation and communication
constraints are linear; hence, \eqref{eq:comm-opt} is an SOCP.

\subsection{Monotonicity with communication availability} \label{subsec:monotonicity}
The communication masks naturally order the admissible controller classes. For two masks $M^{(1)}$ and $M^{(2)}$, define 
\begin{gather*}
    M^{(1)}\preceq M^{(2)}
    \quad\Longleftrightarrow\quad
    m_{ij}^{(1)}\leqslant m_{ij}^{(2)},
    \qquad \forall i,j.
\end{gather*}
Thus, every load-to-generator information link available under $M^{(1)}$ is also available under $M^{(2)}$.
\begin{proposition}[Communication monotonicity]
\label{prop:mon}
For fixed $K_v$,
\begin{gather*}
    M^{(1)}\preceq M^{(2)} \quad\Longrightarrow\quad
    \rho^\star(K_v,M^{(1)}) \leqslant \rho^\star(K_v,M^{(2)}).
\end{gather*}
\end{proposition}
\begin{proof}
If $M^{(1)}\preceq M^{(2)}$, every sparsity constraint imposed by $M^{(2)}$ is also satisfied by any solution feasible under $M^{(1)}$; equivalently, enlarging the mask only relaxes the constraints in \eqref{eq:comm-soc-sparsity}. All other constraints in
\eqref{eq:comm-opt} are unchanged. Hence, the feasible set under $M^{(1)}$ is contained in that under $M^{(2)}$, and the optimal flexibility margin cannot decrease.
\end{proof}
For the $r$-hop masks in \eqref{eq:r-hop-mask}, $r_1\leqslant r_2$ implies
$M^{(r_1)}\preceq M^{(r_2)}$ and therefore
\begin{gather*}
    \rho^\star(K_v,M^{(r_1)}) \leqslant \rho^\star(K_v,M^{(r_2)}).
\end{gather*}

\section{Computational Experiments} \label{sec:results}
This section presents numerical studies on the communication--flexibility trade-off, the impact of available control authority, and non-linear validation of the proposed framework. We first quantify the flexibility recovered as the communication architecture is enriched, then examine the effects of DER capacity and Volt/VAR gain on the achievable margin. Finally, we assess the accuracy of the LinDistFlow-based flexibility certificate against non-linear DistFlow simulations.

\subsection{Test feeder and experimental setup} \label{subsec:data}
The computational experiments are conducted on Feeder~3 of the Iowa-$240$ distribution test system \cite{bu2019time}, modeled as a radial $13.8$~\si{\kilo\volt} feeder. The feeder consists of $162$ buses and $135$ loads downstream of the substation and includes one capacitor location. The feeder is equipped with $5$ DERs, two of them are uncontrollable and inject a fixed value of real and reactive power at a constant power factor and the other $3$ DERs are controllable and shed the active power, if required. The existing capacitor is replaced by a Volt/VAR-controlled device with a nominal reactive-power injection of $40$~\si{\kilo VAR}, a maximum reactive-power capability of $90$~\si{\kilo VAR}, a Volt/VAR gain of $0.4$, and a reference voltage of $1.0$~p.u. Finally for the interchange limits at the feeder head are set to $r_0^\uparrow = 2 \times (\bm 1^\intercal \bar{\bm p}^d)$ and $r_0^\downarrow = 0$. 
All optimization models are implemented in Python using PyOptInterface \cite{yang2024accelerating}. 
The data that includes the parameter values not provided in this paragraph and the source code used to generate the computational results are publicly available at \url{https://github.com/kaarthiksundar/PowerDistribution-Robustness}.

\subsection{Communication--flexibility trade-off} \label{subsec:comm-results}
We first examine how the achievable flexibility changes as the communication architecture is progressively enriched. Recall from Remark~\ref{rem:interpretaion} that the radius $\rho$ measures the Euclidean magnitude of the vector of bus-wise fractional load variations and therefore does not directly represent a percentage increase in feeder demand. We therefore report the flexibility margin in terms of the equivalent uniform load-growth level 
\begin{gather} 
    \lambda^\star(M) \triangleq \rho^\star(K_v,M) \cdot \dfrac{\|\bar{\bm p}^d\|_2}{\|\bar{\bm p}^d\|_1}, \label{eq:load-growth-margin} 
\end{gather} 
which is expressed as a percentage when multiplied by $100$. Thus, $\lambda^\star(M)$ is the uniform fractional increase in all nominal loads whose disturbance vector lies on the boundary of the certified Euclidean uncertainty set. We use $\lambda^\star$ rather than $\rho^\star$ when reporting absolute flexibility margins because it provides a more directly interpretable measure of load-growth capability, while the underlying certificate continues to apply to all, including non-uniform, disturbances contained in the uncertainty set.

To characterize the communication--flexibility trade-off, we normalize both the communication requirement and the resulting flexibility with respect to the full-information architecture. For a communication mask $M$, define the communication density as
\begin{gather} 
    c(M) \triangleq \frac{\|M\|_0}{\|M^{(\infty)}\|_0}, \label{eq:communication-density} 
\end{gather} 
and the corresponding flexibility-recovery ratio 
\begin{gather} 
    f(M) \triangleq \frac{\lambda^\star(M)} {\lambda^\star(M^{(\infty)})} = \frac{\rho^\star(K_v,M)} {\rho^\star(K_v,M^{(\infty)})}. \label{eq:flexibility-recovery} 
\end{gather} 
Thus, $c(M)$ measures the fraction of communication links available relative to the full-information architecture, while $f(M)$ measures the corresponding fraction of full-information flexibility that is recovered. We use communication density, rather than the hop radius $r$, as the primary horizontal-axis variable because successive increases in $r$ can introduce substantially different numbers of communication links. The resulting normalization provides a common basis for quantifying how efficiently additional communication availability translates into flexibility.

\begin{figure}[t]
\centering
\includegraphics[scale=0.7]{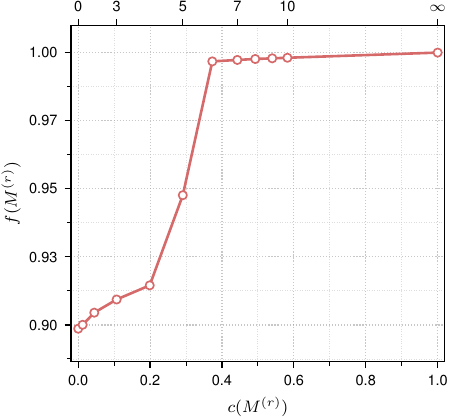}
\caption{Communication--flexibility trade-off for the $r$-hop information architectures. The bottom horizontal axis shows the normalized communication density $c(M^{(r)})$, and the top axis indicates the corresponding hop radius $r$ for selected $r$-hop masks. The vertical axis shows the flexibility-recovery ratio $f(M^{(r)})$ relative to the full-information architecture. For $r=0$, the reported flexibility corresponds to the baseline fixed-controller case obtained by solving the linear program in~\eqref{eq:fixed-controller-opt}; for $r\geqslant 1$, the structured feedforward controller is optimized using~\eqref{eq:comm-opt}.}
\label{fig:comm}
\end{figure}

Figure~~\ref{fig:comm} reveals a clear knee in the communication--flexibility trade-off. Highly localized information structures already recover a substantial fraction of the full-information flexibility, while increasing the communication radius initially yields appreciable additional gains. At a communication density of approximately $0.4$, more than $99\%$ of the full-information flexibility is recovered, after which the curve nearly saturates and additional communication produces only small improvements. Consequently, near-full flexibility can be achieved without requiring the full-information communication architecture. This observation complements Proposition~\ref{prop:mon}, i.e., although the proposition establishes that flexibility is non-decreasing as communication availability increases, Figure~~\ref{fig:comm} shows that the benefit of additional information becomes small once the communication architecture is sufficiently rich.

\subsection{Impact of available control authority} 
We next examine how available DER capacity and local Volt/VAR control affect the flexibility margin. The communication architecture is fixed at the $r=6$ hop configuration, based on the trade-off characterized in the preceding subsection. Starting from the baseline DER configuration, we increase controllable DER capacity by $20\%$-$100\%$ relative to baseline and evaluate Volt/VAR gains $k_v\in\{0.3,0.4,0.5\}$. For each setting, the flexibility margin is recomputed to quantify how additional control authority translates into increased robust demand-growth capability. The baseline case uses $k_v=0.4$.

\begin{figure}
    \centering
    \includegraphics[scale=0.8]{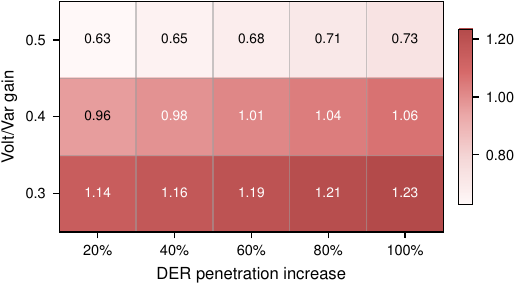}
    \caption{Equivalent uniform load-growth flexibility margin $\lambda^\star(M^{(6)})$ as a function of DER capacity increase relative to the baseline configuration and Volt/VAR gain, with the communication architecture fixed at $r=6$ hops.}
    \label{fig:control-authority}
\end{figure}

The Figure \ref{fig:control-authority} shows that, with the communication architecture fixed at $M^{(6)}$, the flexibility margin increases monotonically with DER capacity for all three Volt/VAR gains. Increasing DER capacity from $20\%$ to $100\%$ above baseline raises $\lambda^\star(M^{(6)})$ from $1.14$ to $1.23$ for $k_v=0.3$, from $0.96$ to $1.06$ for $k_v=0.4$, and from $0.63$ to $0.73$ for $k_v=0.5$. More notably, lower Volt/VAR gains yield substantially larger flexibility margins across the entire DER-capacity range. Thus, in this operating regime, more aggressive local voltage feedback does not necessarily translate into greater system-level flexibility; instead, the choice of $k_v$ can materially alter how effectively the available DER capacity can be used before network or actuator limits become active.  This illustrates a key strength of the proposed framework: it provides a common, certified flexibility metric for quantifying the coupled effects of resource capacity, controller parameters, and communication architecture, rather than assessing each design choice in isolation. The resulting $\lambda^\star$ retains the interpretation of an equivalent uniform load-growth capability while certifying all disturbances contained in the prescribed uncertainty set.

\subsection{DistFlow validation study}
For the non-linear validation, we fix the communication architecture to $M^{(6)}$ and, for brevity, write $\rho^\star$ in place of $\rho^\star(K_v,M^{(6)})$. We evaluate the fixed controller along non-negative disturbance directions $\bm d\geqslant 0$, $\|\bm d\|_2=1$, parameterized as
\begin{gather}
    \bm\eta=\rho\bm d=\gamma\rho^\star\bm d,
    \qquad
    \gamma\triangleq\frac{\rho}{\rho^\star}.
    \label{eq:nonlinear_disturbance}
\end{gather}
Thus, $\gamma$ measures disturbance magnitude relative to the certified LinDistFlow radius, i.e., $\gamma<1$, $\gamma=1$, and $\gamma>1$ correspond to disturbances inside, on, and beyond the certified uncertainty set, respectively. For each constrained quantity $y_i\in[y_i^{\min},y_i^{\max}]$, $i\in\mathcal I$, define the normalized utilization
\begin{gather}
    \chi_i(\bm\eta)\triangleq
    \max\left\{
    \frac{y_i(\bm\eta)-\bar y_i}{y_i^{\max}-\bar y_i},
    \frac{\bar y_i-y_i(\bm\eta)}{\bar y_i-y_i^{\min}}
    \right\},
    \qquad i\in\mathcal I,
    \label{eq:constraint_utilization}
    \\
    \chi(\bm\eta)\triangleq
    \max_{i\in\mathcal I}\chi_i(\bm\eta).
    \label{eq:max_constraint_utilization}
\end{gather}
Hence, $\chi(\bm\eta)$ is the maximum fraction of nominal-to-limit headroom consumed across all engineering constraints; $\chi(\bm\eta)=1$ denotes an active limit and $\chi(\bm\eta)>1$ a violation.

\begin{table}[t]
\caption{Non-linear DistFlow validation. Feas. denotes the number of non-linear-feasible samples, $\max|\Delta V|$ the maximum bus-voltage magnitude error, $\max\|\Delta S\|$ the maximum branch-flow error with $\Delta S=[\Delta P,\Delta Q]^\top$, and $\chi(\bm\eta)$ the maximum non-linear constraint utilization.}
\label{tab:non-linear-validation}
\centering
\setlength{\tabcolsep}{4pt}
\begin{tabular}{ccccc}
\toprule
$\gamma$ & Feas. & $\max |\Delta V|$ & $\max \|\Delta S\|$ & $\chi(\bm{\eta})$ \\
\midrule
0.50 & 212/212 & $1.45\times 10^{-3}$ & $9.01\times 10^{-2}$ & 0.5342 \\
0.75 & 212/212 & $2.03\times 10^{-3}$ & $1.24\times 10^{-1}$ & 0.8064 \\
1.00 & 200/212 & $2.75\times 10^{-3}$ & $1.65\times 10^{-1}$ & 1.0823 \\
1.10 & 200/212 & $3.09\times 10^{-3}$ & $1.84\times 10^{-1}$ & 1.1938 \\
1.25 & 200/212 & $3.64\times 10^{-3}$ & $2.13\times 10^{-1}$ & 1.3622 \\
\bottomrule
\end{tabular}
\end{table}

\begin{figure}
    \centering
    \includegraphics[scale=0.7]{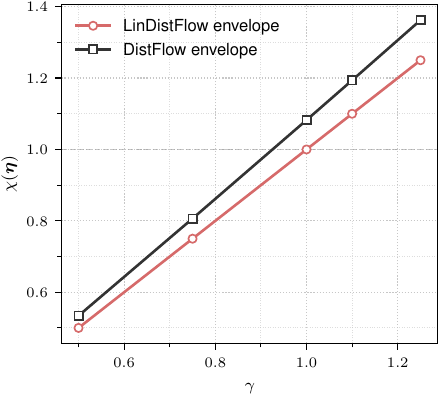}
    \caption{Maximum normalized constraint utilization $\chi(\bm\eta)$ versus normalized disturbance magnitude $\gamma=\rho/\rho^\star$ under LinDistFlow and non-linear DistFlow. The lines $\gamma=1$ and $\chi(\bm\eta)=1$ denote the certified disturbance radius and the feasibility boundary, respectively.}
    \label{fig:non-linear}
\end{figure}

Table~\ref{tab:non-linear-validation} shows that all 212 disturbances remain feasible under DistFlow for $\gamma=0.5$ and $0.75$, with small voltage and branch-flow discrepancies. At the certified boundary, $\gamma=1$, 200 of 212 samples remain feasible and the maximum utilization is $1.0823$, indicating modest optimism of the LinDistFlow certificate near the boundary. The state approximation nevertheless remains accurate as the disturbance grows; even at $\gamma=1.25$, the maximum voltage and branch-flow errors are $3.64\times10^{-3}$ and $2.13\times10^{-1}$, respectively.

Figure~\ref{fig:non-linear} further shows that the non-linear utilization envelope closely tracks the LinDistFlow envelope, but is consistently higher. In particular, the non-linear model reaches $\chi(\bm\eta)=1$ slightly before the LinDistFlow boundary at $\gamma=1$. The close, nearly parallel envelopes indicate that LinDistFlow captures both the location and the growth of network stress with increasing disturbance magnitude, while non-linear effects introduce only a modest shift in the feasibility boundary. Thus, the proposed flexibility margin provides a tractable and interpretable proxy for non-linear system flexibility while making the approximation error explicit.

\section{Conclusion} \label{sec:conclusion}
This paper developed a controller- and communication-aware framework for quantifying operational flexibility in radial distribution networks. Flexibility was defined as the largest prescribed uncertainty set that can be accommodated while satisfying network and actuator limits under a specified control architecture. Under LinDistFlow, the combination of active-power feedforward control and local Volt/VAR feedback yields an affine closed-loop disturbance response, which enables exact robust counterparts of the engineering constraints. For fixed Volt/VAR gains, both global participation-factor design and communication-constrained load-to-generator feedforward control admit convex reformulations, providing a tractable means of co-designing disturbance accommodation and control allocation. 

The numerical results demonstrate that the flexibility margin provides a common quantitative measure for comparing communication structure, DER capability, and controller tuning. Near-full flexibility is recovered with substantially less than full communication, while increased DER capacity and Volt/VAR gain can have markedly different impacts on the achievable margin. Non-linear DistFlow validation further shows that the LinDistFlow-based margin closely tracks the growth of network stress and the non-linear feasibility boundary, with only a modest shift near the certified limit. Collectively, these results show that operational flexibility is determined not only by physical network headroom, but also by how effectively control authority and information are organized. The proposed framework therefore provides a systematic basis for evaluating and designing distribution-system control architectures under uncertainty.

\bibliographystyle{plain}
\bibliography{refs}

\appendix
\subsection{Proof of Proposition \ref{prop:Acl}} \label{subsec:app1}
\begin{proof}
Suppose that $A^{\mathrm{cl}}(K_v)\bm z=\bm 0$. From the block structure of $A^{\mathrm{cl}}(K_v)$,
\begin{gather*}
F\bm p = \bm 0, \quad
F\bm q = K_v\bm v, \quad 2R\bm p+2X\bm q+F^\intercal\bm v = \bm 0,\\
p_0^{\mathrm{sl}} = \bm s^\intercal\bm p, \quad 
q_0^{\mathrm{sl}} = \bm s^\intercal\bm q.
\end{gather*}
Furthermore, since the distribution network is radial, we know that $F$ is nonsingular and we obtain $\bm p=\bm 0$, and hence $p_0^{\mathrm{sl}}=0$. The remaining network equations yield
\begin{gather*}
F\bm q=K_v\bm v, \qquad F^\intercal\bm v=-2X\bm q.
\end{gather*}
Therefore,
\begin{gather*}
\bm v^\intercal K_v\bm v
=\bm v^\intercal F\bm q
=(F^\intercal\bm v)^\intercal\bm q
=-2\bm q^\intercal X\bm q,
\end{gather*}
so that $\bm v^\intercal K_v\bm v+2\bm q^\intercal X\bm q=0$.  Since $K_v\succeq0$ and $X\succeq0$, both terms are nonnegative. Thus $K_v\bm v=\bm 0$, which implies $F\bm q=\bm 0$ and hence $\bm q=\bm 0$. Consequently, $F^\intercal\bm v=\bm 0$, so $\bm v=\bm 0$, and
$q_0^{\mathrm{sl}}=\bm s^\intercal\bm q=0$. Hence $\bm z=\bm 0$, so $A^{\mathrm{cl}}(K_v)$ has a trivial null space and is therefore non-singular.
\end{proof}
\begin{corollary}[Non-singularity of $A$] \label{cor:A-openloop}
The open-loop algebraic operator $A$ in \eqref{eq:open-loop} is non-singular.
\end{corollary}
\begin{proof}
    The result follows since $A^{\mathrm{cl}}(0)=A$.
\end{proof}

\subsection{Proof of Proposition \ref{prop:soc}} \label{subsec:soc-proof}
\begin{proof}
For $\rho>0$, define
\begin{gather*}
    \bm\beta\triangleq\rho\bm\alpha,\qquad
    \beta_0\triangleq\rho\alpha_0,\qquad
    \hat{\Psi}_{\mathrm{cl}} \triangleq\rho\Psi_{\mathrm{cl}}(K_v,\bm\alpha).
\end{gather*}
This change of variables is bijective, with
\begin{gather*}
    \bm\alpha=\frac{\bm\beta}{\rho},\qquad
    \alpha_0=\frac{\beta_0}{\rho},\qquad
    \Psi_{\mathrm{cl}}
    =\frac{\hat{\Psi}_{\mathrm{cl}}}{\rho}.
\end{gather*}
Hence, the participation constraints in \eqref{eq:alpha-policy} are
equivalent to
\begin{gather*}
    \beta_0\geqslant0,\qquad
    \bm\beta\geqslant\bm0,\qquad
    \beta_0+\bm1^\intercal\bm\beta=\rho,
\end{gather*}
together with $\beta_j=0$ at buses without participating active-power
resources. Multiplying \eqref{eq:disturbance-response} by $\rho$ results in
\begin{gather*}
    A^{\mathrm{cl}}(K_v)\hat{\Psi}_{\mathrm{cl}}
    = \rho G_0 - B_p\bm\beta(\bar{\bm p}^d)^\intercal,
\end{gather*}
which is \eqref{eq:disturbance-response-transformed}.

For each constrained quantity, $\hat{\bm\psi}_i=\rho\bm\psi_i(\bm\alpha)$. Since $\rho>0$,
\begin{gather*}
    [\hat{\bm\psi}_i]_+ = \rho[\bm\psi_i(\bm\alpha)]_+, \qquad
    [\hat{\bm\psi}_i]_- = \rho[\bm\psi_i(\bm\alpha)]_-.
\end{gather*}
Therefore, the robust constraints in \eqref{eq:fixed-controller-opt} are equivalent to
\begin{gather*}
    \|[\hat{\bm\psi}_i]_+\|_2 \leqslant
    y_i^{\max}-\bar y_i, \qquad
    \|[\hat{\bm\psi}_i]_-\|_2 \leqslant \bar y_i-y_i^{\min}.
\end{gather*}
Moreover,
\begin{gather*}
    \|[\hat{\bm\psi}_i]_+\|_2\leqslant c
\end{gather*}
holds if and only if there exists $\bm s_i^+$ such that $\bm s_i^+\geqslant\hat{\bm\psi}_i$, $\bm s_i^+\geqslant\bm0$, and $\|\bm s_i^+\|_2\leqslant c$; the corresponding statement holds for the negative part using $\bm s_i^-$. Thus, \eqref{eq:scaled-limits-soc} is an exact representation of the robust constraints, establishing the equivalence with \eqref{eq:vary-alpha-opt} for $\rho>0$.

For $\rho=0$, non-negativity and $\beta_0+\bm1^\intercal\bm\beta=0$ imply
$\beta_0=0$ and $\bm\beta=\bm0$. Then \eqref{eq:disturbance-response-transformed}, together with the non-singularity of $A^{\mathrm{cl}}(K_v)$, yields $\hat{\Psi}_{\mathrm{cl}}=0$. Hence, under the assumed nominal feasibility, $\rho=0$ is feasible and recovery of the participation factors is unnecessary.

Finally, for fixed $K_v$, all constraints in \eqref{eq:vary-alpha-opt} are affine except for the Euclidean-norm bounds in \eqref{eq:participation-soc-upper} and \eqref{eq:participation-soc-lower}, which are second-order-cone constraints. Therefore, \eqref{eq:vary-alpha-opt} is an SOCP.
\end{proof}
\end{document}